\documentstyle[12pt]{article}

\begin{document}
\begin{titlepage}

\begin{flushright}
QMW-PH-94-10\\
{\bf hep-th/9404035}\\
April $6^{th}$, 1994
\end{flushright}

\begin{center}

\baselineskip25pt
{\LARGE {\bf SL(2,R)-DUALITY COVARIANCE OF KILLING SPINORS IN
AXION-DILATON BLACK HOLES}}

\vspace{1cm}

{\large {\bf Tom\'as Ort\'{\i}n}
\footnote{E-mail address: {\tt ortin@qmchep.cern.ch}}\\
{\it Department of Physics}\\
{\it Queen Mary and Westfield College}\\
{\it Mile End Road, London E1 4NS, U.K.}}

\end{center}

\vspace{1.5cm}


\begin{abstract}

Under $SL(2,R)$ electric-magnetic duality transformations the Bogomolnyi
bound of dilaton-axion black holes is known to be invariant.  In this
paper we show that this invariance corresponds to the covariance of the
$N=4$ supersymmetry transformation rules and their parameters.  In
particular this implies that Killing spinors transform covariantly into
Killing spinors.  As an example, we work out completely the case of the
largest known family of axion-dilaton black holes which is
$SL(2,R)$-invariant, finding the Killing spinors with the announced
properties.

\end{abstract}

\end{titlepage}

\newpage

\baselineskip15pt
\pagestyle{plain}


\section*{Introduction}

The SL(2,R)-duality invariance of the low-energy effective string theory
equations of motion \cite{kn:STW,kn:Sendual,kn:Schdual} ($N=4$
supergravity \cite{kn:oldual}) is a fascinating symmetry that
interchanges the strong and weak coupling regimes.  Quantum effects
(instantons \cite{kn:STW} or dyon charge quantization
\cite{kn:Senquan,kn:KO}) effectively break this symmetry to $SL(2,Z)$.

It was conjectured in \cite{kn:firstdual} and more recently in
\cite{kn:Sendual} that $SL(2,Z)$ could be an exact, non-perturbative,
symmetry of superstring theory.  This idea is supported by the fact that
the spectrum of dyon charges \cite{kn:Senquan} is $SL(2,Z)$-invariant.
It is also extremely interesting the interplay between the
$SL(2,Z)$-duality symmetry and other symmetries of compactified string
theory: the $O(6,22;Z)$ symmetry, target-space duality and the
string-fivebrane duality that has been explored in \cite{kn:SchSen} and
\cite{kn:Binet}.

In this letter we are going study the behavior of some supersymmetry
properties under general $SL(2,R)$-duality transformations.  In
particular we will show in Sec. 1 the covariance of the $N=4$
supergravity transformation laws of the gravitino and dilatino fields
under $SL(2,R)$-duality transformations.  As a consequence, Killing
spinors transform covariantly and the number of unbroken supersymmetries
of $SL(2,R)$-related on-shell field configurations is invariant.
Furthermore, the positivity bounds associated to the supersymmetry
algebra are also invariant.  Some of these results were hinted or proven
in special cases in Ref. \cite{kn:TOM} and in a different context in
\cite{kn:Senbogo}.

In Sec. 2 we study the particular case of the $SL(2,R)$-invariant family
of dilaton-axion black holes recently found in Ref. \cite{kn:KO}.  We
find the Killing spinors and check that their behavior under duality
agrees with the results of Sec. 1. The invariance of the
Bogomolnyi-Gibbons-Hull-type bound for axion-dilaton black holes is
built in the solutions and it is, anyway, a straightforward
generalization of the results of Ref. \cite{kn:TOM}.  Therefore we will
nos discuss it.

Our conclusions are in the last section.

Throughout all this paper we use the conventions and notations of Refs.
\cite{kn:KLOPP,kn:KO}, but our definitions of the charges differ
slightly.  The Appendix contains, for the sake of completeness, a
description of the axion-dilaton black holes and the definitions of
charges we are using.


\section{SL(2,R)-covariant Killing spinors}

In the $SU(4)$ version of $N=4$ supergravity, the bosonic part of the
supersymmetry rules of the gravitino and dilatino fields are
respectively \cite{kn:oldual}
\begin{eqnarray}
{\textstyle\frac{1}{2}}\delta_{\epsilon}\Psi_{\mu I} & = &
\nabla_{\mu}\epsilon_{I}
-{\textstyle\frac{i}{4}}e^{2\phi}\partial_{\mu}\epsilon_{I}
-{\textstyle\frac{1}{8}}\sigma^{\rho\sigma}
T_{\rho\sigma\, IJ}^{+}\gamma_{\rho}\epsilon^{J}\; ,
\label{eq:susytrans1}
\\
{\textstyle\frac{1}{2}}\delta_{\epsilon}\Lambda_{I} & = &
-{\textstyle\frac{i}{2}}(e^{2\phi}\not\!\partial\lambda)\epsilon_{I}
+{\textstyle\frac{1}{4}}\sigma^{\rho\sigma}T_{\rho\sigma\,
IJ}^{-}\epsilon^{J}\; ,
\label{eq:susytrans2}
\end{eqnarray}
where
\begin{equation}
\lambda = a+ie^{-2\phi}\, ,
\end{equation}
is a complex scalar field built out of $\phi$, the dilaton field,
and $a$, the axion field. Also
\begin{equation}
T_{\rho\sigma\, IJ} =
{\textstyle2\sqrt{2}}e^{-\phi}[F_{\rho\sigma}\alpha_{IJ}
+iG_{\rho\sigma}\beta_{IJ}]\, ,
\end{equation}
where
$F_{\rho\sigma}=\partial_{\rho}A_{\sigma}-\partial_{\sigma}A_{\rho}$ and
$G_{\rho\sigma}=\partial_{\rho}B_{\sigma}-\partial_{\sigma}B_{\rho}$ are
the field strength tensors of two $U(1)$ vector fields $A_{\mu}$ and
$B_{\mu}$.  Finally, $\alpha_{IJ}$ and $\beta_{IJ}$ are two $SO(4)$
matrices with $SO(4)$ indices $I,J$.

Given any $SL(2,R)$ matrix $\left(\begin{array}{cc}\alpha & \beta \\
\gamma & \delta \\ \end{array} \right)$, $\alpha\delta-\beta\gamma=1$
and the complex scalar field $\lambda$ of the original field
configuration we can define two functions
\begin{eqnarray}
R & = & \alpha\lambda + \beta\, , \nonumber \\
S & = & \gamma\lambda + \delta\, .
\end{eqnarray}
In terms of $R$ and $S$ and the old field con\-fi\-gu\-rations, the
transformed fields can be conveniently written in this way:
\begin{eqnarray}
\lambda^{\prime} & = & R/S\, , \nonumber \\
F^{\pm\prime}_{\rho\sigma} & = & S F^{\pm}_{\rho\sigma}\, ,
\end{eqnarray}
and the analogous equation for $G$. In the Einstein frame (which we are
using) $SL(2,R)$ does not act on the metric.

These transformations rotate continuously the ``Maxwell law" into the
Bianchi identity, hence the name electric-magnetic duality.

It is a well known fact (Refs. \cite{kn:oldual,kn:Sendual}) that the
equations of motion of this theory are invariant under the above
transformations while the action is not.

In addition to this, it is easy to see that the equations
(\ref{eq:susytrans1}) and (\ref{eq:susytrans2}) conserve their form
(i.e. transform covariantly under $SL(2,R)$) if we assume the following
transformation laws for the supersymmetry parameters and variations
\begin{eqnarray}
\epsilon_{I}^{\prime} & = & e^{\frac{i}{2}Arg(S)}\epsilon_{I}\, ,
\label{eq:epsilon}
\\
(\delta_{\epsilon}\Psi_{\mu I})^{\prime} & = &
e^{\frac{i}{2}Arg(S)} \delta_{\epsilon}\Psi_{\mu I}\, ,
\\
(\delta_{\epsilon}\Lambda_{I})^{\prime} & = &
e^{\frac{-3i}{2}Arg(S)} \delta_{\epsilon}\Lambda_{I}\, .
\end{eqnarray}
There is no need to use the equations of motion to prove this result
and, thus, it holds for any on- or off-shell field configuration.

Although this result may not be too surprising we would like to stress
that it is far from being trivial. $SL(2,R)$ is only an invariance of
the classical equations of motion and so it transforms on-shell
configurations into on-shell configurations. There is no apparent
reason why arbitrary field configurations should behave nicely under
$SL(2,R)$. Furthermore, the supersymmetry parameter $\epsilon$ is not
even a field of the theory.

Observe that the transformation (\ref{eq:epsilon}) is consistent with
the invariance of the metric and (accordingly) of the Killing vector
$\xi^{\mu}$ that can be built out of a (commuting) set of Killing
spinors $\epsilon^{I}$:
\begin{equation}
\xi^{\mu}=\overline{\epsilon}^{I}\gamma^{\mu}\epsilon_{I}\, .
\end{equation}

Now let us study some consequences of this result.

Let us assume that for a particular field configuration the equations
$\delta_{\epsilon}\Psi_{\mu I}= \delta_{\epsilon}\Lambda_{I}=0$ are
satisfied by some set of spinors $\epsilon_{I}$, some (perhaps all)
of them trivial (i.e. vanishing). We call this set of spinors ``Killing
spinors". Then the above result implies that in
the duality-transformed field configurations these equations are also
satisfied by another set of Killing spinors, the number of trivial (i.e.
vanishing) and non-trivial Killing spinors being the same as in the
original field configuration.

On-shell configurations admitting non-trivial asymptotically constant
Ki\-lling spinors are said to have unbroken supersymmetries.  Hence, the
number of unbroken supersymmetries is the same for any two
$SL(2,R)$-related classical solutions.

For off-shell configurations the situation is subtly different. They
obey equations of the form
\begin{eqnarray}
\frac{\delta S_{Class}}{\delta g^{\mu\nu}} & = & J_{\mu\nu}\, ,
\nonumber \\
\frac{\delta S_{Class}}{\delta A_{\mu}} & = & J_{A}{}^{\mu}\, ,
\nonumber \\
\frac{\delta S_{Class}}{\delta B_{\mu}} & = & J_{B}{}^{\mu}\, ,
\nonumber \\
\frac{\delta S_{Class}}{\delta \lambda} & = & J\, ,
\label{eq:sources}
\end{eqnarray}
where $S_{Class}$ is the classical action
\begin{equation}
S_{Class}=\int d^{4}x\sqrt{-g}\{-R
-2\frac{\partial_{\mu}\lambda
\partial^{\mu}\overline{\lambda}}{(\lambda-\overline{\lambda})^{2}} +
[i\lambda (F^{+2}+ G^{+2})+c.c.]\}
\end{equation}
and $J_{\mu\nu},J_{A,B}{}^{\mu},J$ are non-vanishing functions
(otherwise the configurations would be on-shell) that may describe
additional matter sources or quantum corrections, for instance.  The
fact that a field configuration admits Killing spinors does not
automatically imply that it obeys the classical equations of
motion\footnote{It does not even imply that we have a ``good" field
configuration.  A set of fields
$\lambda,F_{\mu\nu},G_{\mu\nu},g_{\mu\nu}$ may admit Killing spinors and
still it may not exist a vector fields $A_{\mu},B_{\mu}$ (which are the
true dynamical fields) such that $F_{\mu\nu}=\partial_{\mu} A_{\nu} -
\partial_{\nu} A_{\mu}$ and $G_{\mu\nu}=\partial_{\mu} B_{\nu} -
\partial_{\nu} B_{\mu}$.  See Ref. \cite{kn:Tod} for clear examples.  In
what follows we are going to exclude this possibility, so we will always
have $\nabla^{\mu}{}^{*}F_{\mu\nu}=\nabla^{\mu}{}^{*}G_{\mu\nu}=0$.} It
does not imply, either, that the sources in the right-hand side of eqs.
(\ref{eq:sources}) are coupled in a way consistent with supersymmetry.

In Ref. \cite{kn:KSI} a set of consistency conditions for these sources
to be consistent with supersymmetry was derived. In our case, these
``Killing Spinor Identities" (KSI) take the form
\begin{eqnarray}
\frac{e^{\phi}}{\sqrt{2}}
[J_{A}{}^{\mu}\alpha^{IJ}+i\gamma_{5}J_{B}{}^{\mu}\beta^{IJ}]
\overline{\epsilon}_{J}\gamma_{\mu}
+2iJ\ e^{-2 \phi}\overline{\epsilon}^{I} & = & 0 \, ,
\nonumber\\
2J^{\mu\nu}\overline{\epsilon}^{I}\gamma_{\nu}+
\frac{e^{\phi}}{\sqrt{2}}
[J_{A}{}^{\mu}\alpha^{IJ}+i\gamma_{5}J_{B}{}^{\mu}\beta^{IJ}]
\overline{\epsilon}_{J}
& = & 0 \, ,
\label{eq:ksi}
\end{eqnarray}
where $\epsilon^{I}$ is a set of Killing spinors.

Under an $SL(2,R)$ rotation of the field configuration the sources
transform as follows:
\begin{eqnarray}
(J_{\mu\nu})^{\prime} & = & J_{\mu\nu}\, ,
\nonumber \\
(J_{A,B}{}^{\mu})^{\prime} & = & \alpha J_{A,B}{}^{\mu}\, ,
\nonumber \\
J^{\prime} & = & S^{2}J\, .
\end{eqnarray}
It is obvious that the KSI will generally be violated after a general
$SL(2,R)$ rotation, unless the sources vanish. This result is obviously
related to the fact that a non-vanishing electric current $J$ produces a
non-vanishing magnetic current after the rotation.

Thus, in general, only the unbroken supersymmetries of on-shell field
configurations will be preserved by an electric-magnetic $SL(2,R)$
duality transformation.

This result can be read from a slightly different point of view, as in
Ref. \cite{kn:trace}. If the sources come from quantum corrections and
we select only supersymmetric configurations, then only those with no
quantum corrections will survive. In our case ($N=4$) there is no
trace anomaly and we don't know what the quantum corrections would be
like, but this is clearly a subject worth studying.

Finally, one is tempted to conjecture that in a supersymmetric theory
with manifest $SL(2,R)$ invariant couplings the number of unbroken
supersymmetries would always be invariant, but such a theory does not
yet exist.


\section{The Killing spinors of axion-dilaton black holes}

Recently a quite general family of static black-hole solutions of the
low-energy string theory equations of motion has been presented in
\cite{kn:KO}.  The main feature of this family, which contains many
already known solutions \cite{kn:dilbh,kn:STW,kn:KLOPP}, is that it
constitutes a representation of the whole $SL(2,R)$-duality group in the
sense that by applying any $SL(2,R)$ transformation to any solution in
the family we get another member of the family\footnote{$SL(2,R)$
does not act irreducibly on this family.  In particular, as we are going
to see, this family contains backgrounds with different numbers of
unbroken supersymmetries, and, as a consequence of the results of the
previous section, the corresponding subfamilies provide smaller
representations of the duality group.  Further subdivisions are labeled
by non-duality equivalent values of parameters like the asymptotic value
of asymptotic of $\lambda$.  We won't pursue this issue farther in this
letter.}.

Our immediate goal is to find the non-trivial Killing spinors of
this family of configurations and check explicitly Eq.
(\ref{eq:epsilon}).  We already know that the non-extreme black-hole
solutions in this class do not have non-trivial Killing spinors
\cite{kn:KLOPP}, so we will study only the solutions describing several
extreme black holes in equilibrium, which can be found in the Appendix.

Note that only 6 of the $L$ $U(1)$ fields of the general solution fit
into pure $N=4$ supergravity and we will take only 2 to be non-vanishing
as in the previous section: $A^{(1)}_{\mu}\equiv A_{\mu},\,
A^{(2)}\equiv B_{\mu}$.

We are looking for time-independent ($\partial_{t}\epsilon_{I}=0$)
solutions of the equations
\begin{eqnarray}
\delta_{\epsilon}\Psi_{\mu I} & = 0\, ,
\label{eq:kill1} \\
\delta_{\epsilon}\Lambda_{I} & = & 0\, .
\label{eq:kill2}
\end{eqnarray}
First we contract Eq.(\ref{eq:kill1}) with $\gamma_{\mu}$ and obtain
\begin{equation}
\not\!\nabla\epsilon_{I}-{\textstyle\frac{i}{4}e^{2\phi}\not\!\partial
a\epsilon_{I}}=0\, .
\end{equation}
For time-independent Killing spinors of the above backgrounds this
equation reduces to
\begin{equation}
\gamma^{i}\partial_{\hat{\imath}}
[e^{-\frac{1}{2}(U+i\theta)}\epsilon_{I}] = 0\, ,
\label{eq:kill3}
\end{equation}
where
\begin{equation}
{\cal H}_{2}/\overline{{\cal H}}_{2}=e^{2i\theta}\, ,
\end{equation}
and we have used the property of these solutions
\begin{equation}
\partial_{\hat{\imath}}\ln ({\cal H}_{2}/\overline{{\cal H}}_{2})
= ie^{2\phi}\partial_{\hat{\imath}}a\, .
\end{equation}
If we apply the operator $\gamma^{i}\partial_{\hat{\imath}}$ to Eq.
(\ref{eq:kill3}) we get the integrability condition
\begin{equation}
\partial_{\hat{\imath}}\partial_{\hat{\imath}}
[e^{-\frac{1}{2}(U+i\theta)}\epsilon_{I}] = 0\, ,
\end{equation}
which means that the combination in brackets is a harmonic spinor.
If we substitute now in Eq. (\ref{eq:kill3}) a general harmonic spinor
we see that only a constant one satisfies it, and thus
\begin{equation}
\epsilon_{I}=e^{-\frac{1}{2}(U+i\theta)}\epsilon_{I(0)}\, ,
\label{eq:spinor}
\end{equation}
where the $\epsilon_{I(0)}$s are constant spinors.

This result allows us to check partially the main result in the previous
section. To do this, first observe that under $SL(2,R)$ the functions
${\cal H}_{1}$ and ${\cal H}_{2}$ a transform in this way:
\begin{eqnarray}
{\cal H}_{1}^{\prime} & = & \frac{\alpha {\cal H}_{1} + \beta{\cal
H}_{2}}{\gamma \lambda_{0} + \delta}\, ,
\nonumber \\
{\cal H}_{2}^{\prime} & = & \frac{\gamma {\cal H}_{1} + \delta{\cal
H}_{2}}{\gamma \lambda_{0} + \delta}\, .
\end{eqnarray}
Then, according to Eq. (\ref{eq:spinor})
\begin{eqnarray}
\epsilon_{I}^{\prime} & = & e^{-\frac{1}{2}(U+i\theta^{\prime})}
\epsilon_{I(0)}^{\prime}
\nonumber \\
& = & e^{-\frac{1}{2}U} \left({\cal H}_{2}^{\prime}/\overline{{\cal
H}}_{2}^{\prime}\right)^{-\frac{1}{4}}
\left(\frac{\gamma \lambda_{0} + \delta}{\gamma \overline{\lambda}_{0} +
\delta}\right)^{\frac{1}{4}}\epsilon_{I(0)}^{\prime}
\nonumber \\
& = & e^{\frac{i}{2}Arg(S)}e^{-\frac{1}{2}(U+i\theta)}
e^{\frac{i}{2}Arg(S_{0})}\epsilon_{I(0)}^{\prime}\, ,
\end{eqnarray}
and, looking at Eqs. (\ref{eq:epsilon}) and (\ref{eq:spinor}) we see
that we only need to check that the asymptotic value of the Killing
spinors transforms as follows:
\begin{equation}
\epsilon_{I(0)}^{\prime}= e^{\frac{-i}{2}Arg(S_{0})}\epsilon_{I(0)}\, .
\label{eq:constantspinor}
\end{equation}
where $S_{0}$ is the value of $S$ at infinity.

We only need to find the constant spinors.  It is enough to consider the
equations $\delta_{\epsilon}\Psi_{t I} = \delta_{\epsilon}\Lambda_{I} =
0$. A long but straightforward calculation leads to the following two
equations for each black hole:
\begin{eqnarray}
\frac{ie^{\phi_{0}}}{\sqrt{2}}
[\lambda_{0}M_{i}+\overline{\lambda}_{0}\Upsilon_{i}]
\epsilon_{I(0)}-\{\tilde{Q}_{i}\}_{IJ}\gamma^{0}\epsilon^{J}_{(0)}
& = & 0\, ,
\nonumber \\
\frac{ie^{\phi_{0}}}{\sqrt{2}}
[M_{i}+\Upsilon_{i}]\epsilon_{I(0)} -
\{P_{i}\}_{IJ}\gamma^{0}\epsilon^{J}_{(0)} & = & 0\, ,
\label{eq:constraints}
\end{eqnarray}
where we have used abbreviated notation
\begin{equation}
\{X\}_{IJ}=\{X^{A}\alpha_{IJ}+X^{B}\beta_{IJ}\}\, ,
\end{equation}
for any quantity $X$ associated to the vector fields $A_{\mu}$ and
$B_{\mu}$, the only ones we are considering here.

The algebraic constraints Eqs. (\ref{eq:constraints}) tell us that
the whole solution will have at most as many unbroken supersymmetries as
the single black hole with less unbroken supersymmetries. Note that Eqs.
(\ref{eq:Bogo}), (\ref{eq:charges}) and (\ref{eq:relation}) guarantee
that both
Eqs. (\ref{eq:constraints}) are always compatible for all black holes at
once.

After using the explicit form of the $SO(4)$ matrices $\alpha_{IJ}$ and
$\beta_{IJ}$ (see Ref. \cite{kn:KLOPP}) we can describe the final result
as follows. First we define the central charges
\begin{equation}
z_{\pm}=\sqrt{2}e^{-\phi_{0}}(\Gamma^{A}\pm i\Gamma^{B})\, .
\end{equation}
There are three different cases
\begin{enumerate}
\item One unbroken supersymmetry in the $1,2$ sector with the following
      relations holding for each black hole and the following spinors:
\begin{eqnarray}
M_{i} & = & |z_{-i}|\, ,
\nonumber \\
|\Upsilon_{i}| & = & |z_{+i}|\, ,
\nonumber \\
\mu_{-} & = &
-i\sqrt{2}e^{-\phi_{0}}\frac{\tilde{Q}^{A}
- i\tilde{Q}^{B}}{\lambda_{0} M + \overline{\lambda}_{0}\Upsilon} =
-i\sqrt{2}e^{-\phi_{0}}\frac{P^{A}-iP^{B}}{M + \Upsilon}\, ,
\nonumber \\
\epsilon_{3(0)} & = & \epsilon_{4(0)}=0\, ,
\nonumber \\
\epsilon_{1(0)} & = & \mu_{-i}\gamma^{0}\epsilon^{2}_{(0)}\, .
\end{eqnarray}

\item One unbroken supersymmetry in the $3,4$ sector with the following
      relations holding for each black hole and the following spinors:
\begin{eqnarray}
M_{i} & = & |z_{+i}|\, ,
\nonumber \\
|\Upsilon_{i}| & = & |z_{-i}|\, ,
\nonumber \\
\mu_{+} & = &
-i\sqrt{2}e^{-\phi_{0}}\frac{\tilde{Q}^{A}
+ i\tilde{Q}^{B}}{\lambda_{0} M + \overline{\lambda}_{0}\Upsilon} =
-i\sqrt{2}e^{-\phi_{0}}\frac{P^{A}+iP^{B}}{M + \Upsilon}\, ,
\nonumber \\
\epsilon_{1(0)} & = & \epsilon_{2(0)}=0\, ,
\nonumber \\
\epsilon_{3(0)} & = & \mu_{+i}\gamma^{0}\epsilon^{4}_{(0)}\, .
\end{eqnarray}
\item Two unbroken supersymmetries, one in the $1,2$ sector and one
      in the $3,4$ sector with the following relations holding for
      each black hole and the following spinors:
\begin{eqnarray}
M_{i} & = & |\Upsilon_{i}|=|z_{+i}|=|z_{-i}|\, ,
\nonumber \\
\epsilon_{1(0)} & = & \mu_{-i}\epsilon_{2(0)}\, ,
\nonumber \\
\epsilon_{3(0)} & = & \mu_{+i}\gamma^{0}\epsilon^{4}_{(0)}\, .
\end{eqnarray}

\end{enumerate}

For just one black hole, we always have supersymmetry.  When there is
more than just one, the situation is not so clear.  To have
supersymmetry the complex constants $\mu_{+i}$ or $\mu_{-i}$ must have
the same value for the $N$ black holes.  That this is indeed possible is
clear: just take for one of them a set of charges
$(M_{i},\Sigma_{i},\Delta_{i},Q^{A}_{i},Q^{B}_{i},P^{A}_{i},P^{B}_{i})$
that satisfies the Bogomolnyi-Gibbons-Hull bound Eq.  (\ref{eq:Bogo})
and take for the rest of the black holes sets of charges proportional to
this one. This subclass of configurations has at least one
unbroken supersymmetry. However, due to the large amount of charges and
the involved relations between them we haven't been able to prove that
all of them are supersymmetric\footnote{One might naively think that
it would suffice to use the result of the first section, performing a
duality transformation that would take us to purely electric or
magnetic multi-black-hole solutions. However, these solutions were
found through a duality rotation of multi-black-hole solutions with 2
electric and 2 magnetic charges but no axion. Thus, we could get rid of
the axion charge at most. That simplification is not enough, and
perhaps, a better expression for these solutions is need here.},
although this seems likely.

Back to our main problem, it's easy to see that the constants
$\mu_{\pm}$ transform exactly in the form required by the result of the
previous section, or, equivalently, by Eq.  (\ref{eq:constantspinor})
\begin{equation}
\mu_{\pm}^{\prime} = e^{iArg(S_{0})}\mu_{\pm}\, .
\end{equation}

This illustrates the main result of this paper.

\section{Conclusions}

In this letter we have proven that if a field configuration admits
Killing spinors, then all of its $SL(2,R$) images do, and that the
Killing spinors of the images are related very simply to those of the
original.  This property is related to the $SL(2,R)$ invariance of the
Bogomolnyi-Gibbons-Hull bound of axion-dilaton black holes.  Since this
symmetry does not act on the (Einstein-frame) metric, this result
provides further examples of supersymmetry acting as a cosmic censor,
although one needs to check some consistency conditions (the Killing
Spinor Identities) to see whether supersymmetry is preserved when
quantum corrections are taken into account.

There is a number of other duality symmetries in String Theory, and it
would be interesting to find whether similar properties hold for them.
Target-space duality, for instance, can be seen also as a symmetry of
the low-energy String Theory equations of motion, which can be embedded
in a theory with local supersymmetry. Although it looks very
different from $SL(2,R)$ duality, there are some examples in which
unbroken supersymmetries were preserved by it \cite{kn:BEK}. Note that
$SL(2,R)$ does transform the stringy metric too. We hope to report on
results on this problem soon.


\section*{Acknowledgements}

The author wish to thank R. Kallosh for many discussions and her
support.  This work has been partially supported by a Spanish Government
MEC postdoctoral grant and by a postdoctoral European Communities Human
Capital and Mobility program grant.


\appendix

\section{Axion-dilaton black-hole solutions}

Here we briefly describe the extreme axion-dilaton multi-black-hole
solutions found in Ref. \cite{kn:KO}. Our definitions of the charges
differ slightly from those of that reference.

Apart from the metric and complex scalar field these solutions have $L$
$U(1)$ fields $F^{(n)}_{\mu\nu} = \partial_{\mu} A^{(n)}_{\nu}-
\partial_{\nu} A^{(n)}_{\mu}$.

We also use the auxiliary fields (the $SL(2,R)$-duals)
$\tilde{F}^{(n)}_{\mu\nu} = e^{-2\phi} {}^{*}F^{(n)}_{\mu\nu}- ia
F^{(n)}_{\mu\nu}$.  Owing to the ``Maxwell law", which can be written
$\nabla^{\mu} {}^{*}\tilde{F}_{\mu\nu} = 0$, there exist locally $L$
real vector fields $\tilde{A}^{(n)}_{\nu}$ such that $\tilde{F}_{\mu\nu}
= i(\partial_{\mu} \tilde{A}^{(n)}_{\nu}- \partial_{\nu}
\tilde{A}^{(n)}_{\mu}$).  They simplify the description of the
solutions, which are given by
\begin{eqnarray}
ds^{2} & = &
e^{2U}dt^{2}-e^{-2U}d\vec{x}^{2}\, ,
\nonumber \\
e^{-2U}(\vec{x}) & = &
2 \,{\mbox{Im}}\,
({\cal H}_{1}(\vec{x})\, \overline{{\cal H}}_{2}(\vec{x}))\, ,
\nonumber \\
\lambda(\vec{x}) & = &
\frac{{\cal H}_{1}(\vec{x})}{{\cal H}_{2}(\vec{x})}\, ,
\nonumber \\
A_{t}^{(n)}(\vec{x}) & = &
e^{2U}(k^{(n)}{\cal H}_{2}(\vec{x})+c.c.)\, ,
\nonumber \\
\tilde{A}^{(n)}_{t}(\vec{x}) & = &
-e^{2U}(k^{(n)}{\cal H}_{1}(\vec{x})+c.c.)\, ,
\end{eqnarray}
where ${\cal H}_{1}(\vec{x}),\, {\cal H}_{2}(\vec{x})$
are two complex harmonic functions with $N$ poles corresponding to $N$
labeled by $i=1,\ldots,N$
\begin{eqnarray}
{\cal H}_{1}(\vec{x}) & = &
\frac{e^{\phi_{0}}}{\sqrt{2}} \{\lambda_{0}+\sum_{i=1}^{N}
\frac{\lambda_{0}M_{i}+\overline{\lambda}_{0}
\Upsilon_{i}}{|\vec{x}-\vec{x}_{i}|}\}\, ,
\nonumber \\
{\cal H}_{2}(\vec{x}) & = &
\frac{e^{\phi_{0}}}{\sqrt{2}} \{1+\sum_{i=1}^{N}\frac{M_{i}+
\Upsilon_{i}}{|\vec{x}-\vec{x}_{i}|}\}\, ,
\nonumber \\
\partial_{\hat{\imath}}\partial_{\hat{\imath}}{\cal H}_{1}  & = &
\partial_{\hat{\imath}}\partial_{\hat{\imath}}{\cal H}_{1}  = 0\, .
\end{eqnarray}
The different constants, which are defined later, have to satisfy
several identities. First, they have to satisfy
Bogomolnyi-Gibbons-Hull-type identities for each black hole and for
the global solution:
\begin{eqnarray}
M^{2}_{i}+|\Upsilon_{i}|^{2}+2i(\lambda_{0}-
\overline{\lambda}_{0})\sum_{n=1}^{L}|\Gamma^{(n)}_{i}|^{2} & = & 0\, ,
\nonumber \\
M^{2}+|\Upsilon|^{2}+2i(\lambda_{0}-
\overline{\lambda}_{0})\sum_{n=1}^{L}|\Gamma^{(n)}|^{2} & = & 0\, .
\label{eq:Bogo}
\end{eqnarray}
Secondly, the consistency of the solution requires
\begin{eqnarray}
k^{(n)} & = & -\sqrt{2}e^{-\phi_{0}}
\frac{\Gamma^{(n)}M+
\overline{\Gamma}^{(n)}\overline{\Upsilon}}{M^{2}-|\Upsilon|^{2}}
=k^{(n)}_{i}\, ,
\nonumber \\
Arg(\Upsilon) & = & Arg(\Upsilon_{i})\, ,
\label{eq:charges}
\end{eqnarray}
for each black hole $i$.

Finally, $\Upsilon$ and $\Gamma$ are related by
\begin{equation}
\Upsilon = i(\lambda_{0}-\overline{\lambda}_{0})
\frac{\sum_{n=1}^{L}\overline{\Gamma}^{(n)2}}{M}\, ,
\label{eq:relation}
\end{equation}
for each black hole and for the global solutions.

The different constants are defined by the asymptotic expansions in the
``upper sheet" (global charges) and in the $i^{th}$ black hole
sheet:
\begin{eqnarray}
g_{tt} & \sim & 1-\frac{2M}{r}\, .
\nonumber \\
F^{(n)+} & \sim & \frac{\Gamma^{(n)}}{r^{2}}\, ,
\nonumber \\
\tilde{F}^{(n)+} & \sim & \frac{\tilde{\Gamma}^{(n)}}{r^{2}}\, ,
\nonumber \\
\lambda & \sim & \lambda_{0}-
(\lambda_{0}-\overline{\lambda}_{0})\frac{\Upsilon}{r}\, .
\end{eqnarray}
We can write these complex constants in terms of real constants:
electric ($Q$) and magnetic ($P$) charges,
dilaton ($\Sigma$) and axion ($\Delta$) charges and dilaton $\phi_{0}$
and axion $a_{0}$ asymptotic values:
\begin{eqnarray}
\Gamma^{(n)} & = & \frac{1}{2}(Q^{(n)}+iP^{(n)})\, ,
\nonumber \\
\Upsilon & = & \Delta-i\Sigma\, ,
\nonumber \\
\lambda_{0} & = & a_{0}+ie^{-2\phi_{0}}\, .
\end{eqnarray}
The charges in the upper sheet (the global charges) are the sum of the
charges of the $N$ black holes.


\end{document}